\documentstyle[aps]{revtex}
\begin{document}
\draft
\author{V.B. Svetovoy\thanks{E-mail: svetovoy@nordnet.ru} and M.V. Lokhanin}
\address{Department of Physics, Yaroslavl State University, \\
Sovetskaya 14, Yaroslavl 150000, Russia }
\title{Fulfillment of expectations of precise measurements of the Casimir force }
\date{\today}
\maketitle

\begin{abstract}
We compare theoretical expectations for the Casimir force with the results
of precise measurements. The force is calculated at finite temperature for
multilayered covering of the bodies using the Lifshitz theory. We argue
that the dielectric function of the metallization has to be directly
measured to reach the necessary precision in the force calculation. Without
knowledge of this function one can establish a well defined upper limit on
the force using parameters of perfect single-crystal materials. The force
measured in the torsion pendulum experiment does not contradict to the upper
limit. Importance of a thin $Au/Pd$ layer in the atomic force microscope
experiments is stressed. The force measured with the microscope is larger
than the upper limit at small separations between bodies. The discrepancy is
significant and reproduced for both performed measurements. The origin of
the discrepancy is discussed. The simplest modification of the experiment is
proposed allowing to make its results more reliable and answer the question
if the discrepancy has any relation with the existence of a new force.
\end{abstract}
\pacs{12.20.Ds, 03.70.+k}

\section{Introduction}

The Casimir force \cite{Casimir} (see \cite{book} for a review) between
closely spaced macroscopic bodies is an effect of quantum electrodynamics
(QED) and for this reason it could be predicted very accurately. The force
acting between nonideal bodies can be found using the Lifshitz theory \cite
{Lif,LP}, where it depends on optical properties of used materials.
Knowledge of these properties is the weakest element in the theory
restricting the accuracy that can be achieved. Experiments measuring the
Casimir force are of great importance because they are sensitive to the
presence of new fundamental forces \cite{Kuz,Most1} predicted in many modern
theories (see \cite{Long,Fisch} and references therein). To distinguish a
new force from the background, we should be able to calculate the Casimir
force with a precision better than the experimental one. In the series of
recent experiments this force has been measured with the torsion pendulum
(TP) \cite{Lam1} in the range of distances $0.6-6\ \mu m$ and with the
atomic force microscope (AFM) \cite{MR,RM,RLM} in the range $0.1-0.9\ \mu m$%
. The corresponding precisions were 5\% and 1\%, respectively.

For two ideal plates the famous Casimir formula \cite{Casimir} for the force
per unit area is

\begin{equation}
\label{Fc}F_c^{pl}(a)=\frac{\pi ^2\hbar c}{240a^4}, 
\end{equation}

\noindent where $a$ is the distance between plates. In the experiments the
force is measured between metallized disc and sphere because for two plates
it is difficult to keep them parallel. In this case (\ref{Fc}) has to be
modified with the proximity force theorem (PFT) \cite{PFT}. This theorem
allows to evaluate the force by adding the contributions of various
distances as if they were independent and for plate and sphere it is reduced
to

\begin{equation}
\label{PFT}F(a)=2\pi R\int\limits_a^{R+a}F^{pl}\left( x\right) dx, 
\end{equation}

\noindent where $R$ is the radius of curvature of the spherical surface. The
PFT approximation is good for $R\gg a$ that holds true in all the
experiments. If we use the Casimir expression (\ref{Fc}) for the force in (%
\ref{PFT}), then the force between plate and sphere will be

\begin{equation}
\label{F0}F_c^0(a)=\frac{\pi ^3\hbar c}{360}\frac R{a^3}. 
\end{equation}

Eq.\ (\ref{F0}) was deduced for ideally conducting bodies at zero
temperature and three kinds of corrections have been considered to take into
account their real properties. The correction due to finite metal
conductivity was found \cite{Cond1,Cond2} on the base of the free electron
model, where the optical properties of a metal were described by the only
parameter $\omega _p$ which is the plasma frequency. The force including
corrections up to the second order \cite{Cond3} is

\begin{equation}
\label{Fp}F_c^p\left( a\right) =F_c^0\left( a\right) \left[ 1-4\frac c{%
a\omega _p}+\frac{72}5\left( \frac c{a\omega _p}\right) ^2\right] . 
\end{equation}

\noindent For typical plasma frequency $\omega _p\sim 10^{16}\ s^{-1}$ and
separations $a\leq 1\ \mu m$ the correction will be more than 10\%.
Correction due to finite temperature has been found \cite{Temp} for ideal
conductors and the resulting force is given by

\begin{equation}
\label{FT}F_c^T\left( a\right) =F_c^0\left( a\right) \left( 1+\frac{720}{\pi
^2}f\left( \xi \right) \right) , 
\end{equation}

\noindent where $\xi =k_BTa/\hbar c\approx a(\mu m)/7.61$ for $T=300^{\circ
}K$. The function $f(\xi )$ is expressed via an infinite sum but it can be
represented approximately as $f(\xi )=(\xi ^3/2\pi )\zeta (3)-(\xi ^4\pi
^2/45)$ for $\xi <1/2$. The temperature correction is negligible for the AFM
experiments \cite{MR,RLM} since $\xi $ is small in the important separation
range $0.1-0.3\ \mu m$ and is only a minor correction in condition of the TP
experiment \cite{Lam1}, where the important separation range was $0.6-3\ \mu
m$. The general form of the correction due to surface distortions has been
found in \cite{Dist}. If the bodies are covered by distortions with
characteristic amplitudes $A_1$ and $A_2$, then the force up to the second
order in the relative amplitudes of the distortions has the form

$$
F_c^d\left( a\right) =F_c^0\left( a\right) \left[ 1+3\left( \left\langle
f_1\right\rangle \frac{A_1}a-\left\langle f_2\right\rangle \frac{A_2}a%
\right) +\right. 
$$

\begin{equation}
\label{disflat}\left. 6\left( \left\langle f_1^2\right\rangle \frac{A_1^2}{%
a^2}-2\left\langle f_1f_2\right\rangle \frac{A_1A_2}{a^2}+\left\langle
f_2^2\right\rangle \frac{A_2^2}{a^2}\right) \right] , 
\end{equation}

\noindent where the functions $f_{1,2}\left( x,y\right) $ describe
distribution of the distortions on the surfaces and $\left\langle
...\right\rangle $ denotes averaging over the surface area. Corrections due
to surface roughness are very important for the experiment \cite{MR}.

At first \cite{Lam1,MR,RM,KRMM} the experimental data were treated using
these corrections to Eq.\ (\ref{F0}), but it was realized soon that at least
the conductivity correction has to be considered on more reliable basis. In
more realistic approach the Lifshitz theory \cite{LP} was used to evaluate
the force between bodies \cite{RLM,Lam2}. Similar but technically a little
bit different method was developed in \cite{LR1}. In these approaches the
force depends on the dielectric function of the bodies at imaginary
frequencies $\varepsilon \left( i\omega \right) $. It has to be expressed
with the dispersion relation via the imaginary part of the function $%
\varepsilon \left( \omega \right) $ on the real axis which can be directly
measured. However, in any of the experiments the information on $\varepsilon
\left( \omega \right) $ was not collected and the handbook data were used
instead. Such data are good only to make an estimate for the Casimir force
with the accuracy much worse than the experimental one. The reason is that
the dielectric function depends in substantial degree on the sample
preparation procedure as will be discussed below. Nevertheless, it is
possible to find \cite{SL} a reliable upper limit on the Casimir force using
only well defined parameters of perfect crystalline materials. In this paper
we will discuss in detail this limit and its comparison with the existing
experimental data.

The paper is organized as follows. In Sec.\ \ref{2} we give a general
expression for the Casimir force between sphere and plate made of nonideal
materials at nonzero temperature. Then, to treat the experimental data, the
expression for the force is generalized for the case of layered bodies. The
choice of dielectric functions and parameters for the used materials is
described in Sec.\ \ref{3}. In Sec.\ \ref{4} we define the boundary values
of the optical parameters and find the upper limit on the force in
conditions of each independent experiment. Possible reasons for discrepancy
between theory and experiment are discussed in Sec.\ \ref{5}. Our
conclusions are given in the last Section.

\section{\label{2}Theory}

Let us discuss first a reliable way to evaluate the Casimir force in the
experimental configurations. The force per unit area between parallel plates
arising as a result of electromagnetic fluctuations is generalized by the
Lifshitz theory \cite{LP}, where the real material is taken into account by
its dielectric function at imaginary frequencies $\varepsilon \left( i\zeta
\right) $:

\begin{equation}
\label{Liff}F^{pl}(a)=\frac{kT}{\pi c^3}{\sum\limits_{n=0}^\infty {}}%
^{\prime }\zeta _n^3\int\limits_1^\infty dpp^2\left\{ \left[ G_1^2e^{2p\zeta
_na/c}-1\right] ^{-1}+\left[ G_2^2e^{2p\zeta _na/c}-1\right] ^{-1}\right\} . 
\end{equation}

\noindent Here prime over the sum sign means that $n=0$ term is taken with
the coefficient $1/2$ and

$$
G_1=\frac{p+s}{p-s},\quad G_2=\frac{\varepsilon \left( i\zeta _n\right) p+s}{%
\varepsilon \left( i\zeta _n\right) p-s},\quad 
$$

\begin{equation}
\label{defin1}s=\sqrt{\varepsilon \left( i\zeta _n\right) -1+p^2},\quad
\zeta _n=\frac{2\pi nkT}\hbar . 
\end{equation}

\noindent It is supposed that both bodies were made of identical materials.
The function $\varepsilon \left( i\zeta _n\right) $ cannot be measured
directly but can be expressed via imaginary part of the dielectric function $%
\varepsilon ^{\prime \prime }\left( \omega \right) $ on the real axis with
the dispersion relation

\begin{equation}
\label{disp}\varepsilon \left( i\zeta \right) -1=\frac 2\pi
\int\limits_0^\infty d\omega \frac{\omega \varepsilon ^{\prime \prime
}\left( \omega \right) }{\omega ^2+\zeta ^2}. 
\end{equation}

\noindent Information on $\varepsilon ^{\prime \prime }\left( \omega \right) 
$ can be extracted from the data on reflectivity and absorptivity of
electromagnetic waves with the frequency $\omega $ for a given material.

Applying PFT to Eq.\ (\ref{Liff}) one can find the force between sphere and
plate. The integration in (\ref{PFT}) can be done analytically and we find

\begin{equation}
\label{shpl}F(a)=-\frac{kTR}{c^2}{\sum\limits_{n=0}^\infty {}}^{\prime
}\zeta _n^2\int\limits_1^\infty dpp\ln \left[ \left( G_1^{-2}e^{-2p\zeta
_na/c}-1\right) \left( G_2^{-2}e^{-2p\zeta _na/c}-1\right) \right] . 
\end{equation}

\noindent Special care needs to treat the first $n=0$ term. The formal
reason is that $\zeta _n^2$ becomes zero but the integral over $p$ diverges.
The physical reason is that this term corresponds to the static limit when
for metallic bodies $\varepsilon \rightarrow \infty $. This means that any
parameter characterizing the dielectric function of a metal cannot appear in
the $n=0$ term in contrast with a dielectric for which it will depend on the
static permittivity of the material. In the $\varepsilon \rightarrow \infty $
limit the functions $G_{1,2}$ become $-G_1=G_2=1$. The formal problem is
overcome by introducing the integration over a new variable $x=2p\zeta _na/c$
and after that one can take $\zeta _n=0$ for the $n=0$ term. The resulting
contribution of the first term in the force corresponds to the classical
limit $F_{cl}\left( a\right) $ for metals

\begin{equation}
\label{clas}F_{cl}\left( a\right) =\frac{kTR}{4a^2}\zeta \left( 3\right) , 
\end{equation}

\noindent where $\zeta \left( n\right) $ is the zeta-function. Note that in
this limit the force does not depend on the metal parameters as it should be
for a static field.

The bare Casimir force (\ref{F0}) is reproduced from Eq.\ (\ref{shpl}) in
the limit $\varepsilon \rightarrow \infty $ and $T\rightarrow 0$. The finite
conductivity correction also can be derived from (\ref{shpl}). To this end
one considers the limit of small temperature when the sum in (\ref{shpl})
can be replaced by the integral and supposes that the dielectric function of
the metal covering the bodies is described by the free electron plasma
model. In this model $\varepsilon \left( i\zeta \right) $ is

\begin{equation}
\label{plasma}\varepsilon \left( i\zeta \right) =1+\frac{\omega _p^2}{\zeta
^2}, 
\end{equation}

\noindent where $\omega _p$ is the free electron plasma frequency. Typical
value of the frequency $\omega _p\sim 10^{16}\ s^{-1}$ is larger than
fluctuation frequencies $\zeta \sim c/a$ giving the main contribution in (%
\ref{shpl}). Then one can expand the functions $G_{1,2}$ in (\ref{shpl}) in
powers of the parameter $\zeta /\omega _p$ and performing necessary
integrations one finds exactly the result (\ref{Fp}) for the conductivity
corrections \footnote[1]{%
The correction is actually connected with finite density of free electrons
(finite $\omega _p$) since the metal conductivity is still infinite for the
plasma model. Nevertheless, we will not change the fixed terminology.}. In
this way the corrections up to the fourth order were found in recent paper 
\cite{Cond4}. The temperature correction (\ref{FT}) is also reproduced from (%
\ref{shpl}) in the limit of ideal metals $\varepsilon \rightarrow \infty $.
In this case the linear in temperature correction does not survive since the 
$n=0$ term (\ref{clas}) is exactly canceled by the linear in $T$
contribution from the rest terms in the sum. As a result the leading
correction behaves only as $\xi ^3$.

The expression (\ref{shpl}) differs from those used in \cite{RLM} and \cite
{Lam2} in two respects. First, in the cited papers the integration connected
with the PFT was not done analytically that complicated numerical analysis.
Second, the zero temperature limit has been taken. This limit was also
considered in \cite{LR1}, though the PFT integral was evaluated explicitly.
It seems a reasonable approximation at small separations because the
temperature correction in (\ref{FT}) is proportional to $\xi ^3$ and,
therefore, is small. However, one should remember that this correction was
derived in the limit of ideal conductors $\varepsilon \rightarrow \infty $.
For a real conductor it will be proportional to $\xi $ as expected for
difference between sum and integral and will be important (for details see 
\cite{SLtem}). We have computed the force according to (\ref{shpl}) and with
the integral instead of the sum at the smallest separation $a=100\ nm$
tested in the experiments. For the plasma model (\ref{plasma}) with $\omega
_p=2\cdot 10^{16}\ s^{-1}$ we have found that the difference between the sum
and integral is $2.5\ pN$ for $T=300^{\circ }\ K$. It becomes $4\ pN$ for
the Drude dielectric function (see Eq.\ (\ref{dfimag}) below) with the
damping frequency $\omega _\tau =5\cdot 10^{13}\ s^{-1}$. These values
exceed the conservative estimate for the experimental errors $2\ pN$ \cite
{RLM}.

In the AFM experiments an additional $Au_{0.6}Pd_{0.4}$ layer of $20\ nm$ 
\cite{MR} or $8\ nm$ \cite{RM,RLM} thick was on the top of $Al$
metallization of the bodies to prevent aluminum oxidation. It has to be
included into consideration. This layer is transparent for the
electromagnetic waves with high frequencies $\sim c/a$ since adsorption,
proportional to $\varepsilon ^{\prime \prime }\left( \omega \right) $, is
small. For this reason the layer was ignored in \cite{MR,RM,RLM}. However,
the force depends on $\varepsilon (i\zeta )$ for which the low frequencies
dominate in the dispersion relation (\ref{disp}) because of large $%
\varepsilon ^{\prime \prime }\left( \omega \right) $ and that is why we
cannot neglect the $Au/Pd$ layer. To take it into account, one has to
generalize expression for the force (\ref{Liff}) to the case of layered
bodies. Suppose that the top layer has the thickness $h$ and its dielectric
function is $\varepsilon _1$. The bottom layer is thick enough to be
considered as infinite and let its dielectric function be $\varepsilon _2$.
The method described in \cite{LP} for deriving Eq.\ (\ref{Liff}) can be
easily generalized for layered plates. We have to add only the matching
conditions for the Green functions on the layers interface. After some
algebra the result will look exactly as (\ref{Liff}) but with more complex $%
G_{1,2}$:

$$
G_1=\frac{\left( s_1+s_2\right) \left( p+s_1\right) e^{\zeta
_ns_1h/c}+\left( s_1-s_2\right) \left( p-s_1\right) e^{-\zeta _ns_1h/c}}{%
\left( s_1+s_2\right) \left( p-s_1\right) e^{\zeta _ns_1h/c}+\left(
s_1-s_2\right) \left( p+s_1\right) e^{-\zeta _ns_1h/c}}, 
$$

\begin{equation}
\label{defin2}G_2=-\frac{\left( \varepsilon _2s_1+\varepsilon _1s_2\right)
\left( \varepsilon _1p+s_1\right) e^{\zeta _ns_1h/c}+\left( \varepsilon
_2s_1-\varepsilon _1s_2\right) \left( \varepsilon _1p-s_1\right) e^{-\zeta
_ns_1h/c}}{\left( \varepsilon _2s_1+\varepsilon _1s_2\right) \left(
\varepsilon _1p-s_1\right) e^{\zeta _ns_1h/c}+\left( \varepsilon
_2s_1-\varepsilon _1s_2\right) \left( \varepsilon _1p+s_1\right) e^{-\zeta
_ns_1h/c}}, 
\end{equation}

\noindent where $s_{1,2}$ are defined similar to $s$ in (\ref{defin1}). The
force between plate and sphere is given by (\ref{shpl}) with the above $%
G_{1,2}$.

To see qualitatively the effect of an additional layer, we found the finite
conductivity correction up to the second order in this case. Than for the
force one has

\begin{equation}
\label{Fph}F_c^p\left( a,h\right) =F_c^0\left( a\right) \left[ 1-4K(h)\frac c%
{a\omega _{1p}}+\frac{72}5\left( K(h)\frac c{a\omega _{1p}}\right) ^2\right]
, 
\end{equation}

\noindent where the function $K(h)$ depends on the plasma frequencies of the
layers $\omega _{1p}$ , $\omega _{2p}$ and the thickness of the top layer $h$

\begin{equation}
\label{Ah}K(h)=\frac{\omega _{1p}+\omega _{2p}\tanh \left( h\omega
_{1p}/c\right) }{\omega _{2p}+\omega _{1p}\tanh \left( h\omega
_{1p}/c\right) }. 
\end{equation}

\noindent When $h=0$ the force will depend only on $\omega _{2p}$ and in the
case $h\rightarrow \infty $ on $\omega _{1p}$ as it should be. The effect of
the top layer disappears if the plasma frequencies coincide. The top layer
will be negligible if $h\omega _{1p}/c\ll 1$. For typical plasma frequencies 
$\sim 10^{16}\ s^{-1}$ it is definitely not the case even for $h=8\ nm$. The
opposite conclusion made in \cite{KRMM} was based on the too small value of $%
\omega _p$ for gold as will be discussed below (see also \cite{LR1}). Eq.\ (%
\ref{Fph}) is not very good approximation and was discussed only for
qualitative understanding of the effect. For actual calculations we will use
the exact equations (\ref{shpl}), (\ref{defin2}).

Importance of a thin metallic layer on the body surfaces has been stressed
first in \cite{Lam3}. The general expression for the Casimir force between
layered bodies has been presented in \cite{LR1} but was not used their for
actual calculations. Significant role of the $Au/Pd$ layer in the AFM
experiments was indicated in our preprint \cite{SL}, where it was
demonstrated that the effect far exceeds the experimental errors. This
conclusion was supported in \cite{KMM}, where the expressions (\ref{defin2})
for $G_{1,2}$ were confirmed using a different method to deduce them.
However, the authors were uncertain on applicability of (\ref{defin2}) for
thin films with $h<25\ nm$ because the spatial dispersion of the dielectric
function can be important for such films. We discuss this effect in Sec.\ 
\ref{5} where we argue that the spatial dispersion can be neglected because
of very short mean free path for the electrons in thin films.

\section{\label{3}The dielectric function}

Now we are able to evaluate the Casimir force in real geometry of the
experiments if there is information on the dielectric functions of used
materials: $Au$, $Al$, and $Au_{0.6}Pd_{0.4}$ alloy. Strictly speaking, one
has to measure these functions in wide range of wavelengths on the same
samples which are used for the force measurement. It was not done in all of
the experiments and to draw any conclusion from them we have to make some
assumptions on the dielectric functions. At low frequencies $Au$ and $Al$
are well described by the Drude dielectric function \cite{Gros}: 
$$
\varepsilon =\varepsilon ^{\prime }+i\,\varepsilon ^{\prime \prime }, 
$$

\begin{equation}
\label{Drudr}\varepsilon ^{\prime }\left( \omega \right) =1-\frac{\omega _p^2%
}{\omega ^2+\omega _\tau ^2},\qquad \varepsilon ^{\prime \prime }\left(
\omega \right) =\frac{\omega _p^2\omega _\tau }{\omega \left( \omega
^2+\omega _\tau ^2\right) }, 
\end{equation}

\noindent where $\omega _p$ is the free electron plasma frequency and $%
\omega _\tau $ is the Drude damping frequency. A simple test for validity of
the Drude model is behavior of the material resistivity \cite{Pers} which is
defined as

\begin{equation}
\label{res}\rho \left( \omega \right) =Im\left( \frac 1{\varepsilon _0\left(
1-\varepsilon \left( \omega \right) \right) \omega }\right) =\frac{\omega
_\tau }{\varepsilon _0\omega _p^2}, 
\end{equation}

\noindent where $\varepsilon _0$ is the free space permittivity. The
resistivity is frequency independent within the Drude approximation. For
high-purity single-crystal samples of $Au$ and $Al$ (entries 2 in Table\ \ref
{1t}) the frequency behavior of the resistivity in the infrared range of
wavelengths $3\ \mu m<\lambda <32\ \mu m$ is shown in Fig.\ \ref{1f}. The
data on the dielectric functions were taken from \cite{Zol}, where the
results from many original works are collected. The data for $\varepsilon
^{\prime }\left( \omega \right) $ and $\varepsilon ^{\prime \prime }\left(
\omega \right) $ can be fitted with (\ref{Drudr}) to find the parameters $%
\omega _p$ and $\omega _\tau $. The points and fitting curves for $%
\varepsilon ^{\prime \prime }\left( \omega \right) $ are shown in the same
Fig.\ \ref{1f}. Palladium definitely cannot be described by (\ref{Drudr})
since its resistance significantly changes in the infrared range. However,
it is known experimentally that amorphous metallic alloys can be described
by the Drude approximation \cite{Pers}. The physical explanation for this is
associated with large Drude damping of the compounds like $Au_{0.6}Pd_{0.4}$.

Of course, at higher frequencies when interband transitions are reached the
Drude approximation fails. Nevertheless, it is very helpful since low
frequencies dominate in the dispersion relation. Extrapolating (\ref{Drudr})
to all frequencies one finds

\begin{equation}
\label{dfimag}\varepsilon \left( i\zeta \right) =1+\frac{\omega _p^2}{\zeta
\left( \zeta +\omega _\tau \right) }. 
\end{equation}

\noindent Let us estimate the relative error inserted in (\ref{dfimag}) due
to extrapolation. If $\omega _0$ is the frequency of the first resonance for
a given metal, then the contribution in $\varepsilon \left( i\zeta \right) $
of the frequency range $\omega _0<\omega <\infty $, where the Drude model
does not valid, will be $\left( \omega _p/\zeta \right) ^2$ $\cdot \left(
\omega _\tau /\omega _0\right) $ for $\zeta \geq \omega _0$. This
contribution one can take as an estimate for the absolute error and,
therefore, for the relative error one has $\sim \omega _\tau /\omega _0$.
For typical values $\omega _\tau \sim 10^{14}\ s^{-1}$ and $\omega
_0>10^{15}\ s^{-1}$ the error can be as large as 10\% but error in the force
is smaller. If we will use (\ref{dfimag}) for the force computation and
change $\omega _p$ by 5\% ( 10\% correction to $\varepsilon \left( i\zeta
\right) $ at all frequencies) then the force is changed less than 2\%.
Moreover, since the interband transitions give a correction to (\ref{dfimag}%
) which is frequency dependent, it reduces the correction to the force
further. Of course, we can take the interband transitions into consideration
exactly using the handbook data in visible-ultraviolet range which are not
very sensitive to the purity and defect density as it happens in the
infrared range. However, we are intended to establish the upper limit on the
force using $\omega _p$ in (\ref{dfimag}) which is definitely larger than any
real value and for this reason we can neglect the interband transitions.

Therefore, in all cases of interest we can use Eq.\ (\ref{dfimag}) to
describe the dielectric function of a material on the imaginary axis. The
question is how we should extract the parameters $\omega _p$ and $\omega
_\tau $ from the data. We proceeded as follows. The data for the complex
refraction index $n+i\kappa =\sqrt{\varepsilon }$ have been taken from \cite
{Zol}. First, the validity of the Drude approximation was checked by
calculating the frequency dependence of the material resistivity according
to (\ref{res}). In the investigated cases the resistivity is more or less
constant in the wavelength range $\lambda >2\ \mu m$ ($\omega <9.4\cdot
10^{14}\ s^{-1}\ $). This range gives the most important contribution to the
dispersion relation (\ref{disp}) and, therefore, it is the range where we
have to extract the Drude parameters. Using $\omega _p$ found from the high
frequency region can happen to be wrong. For example, sometime the plasma
frequency is estimated using the transition point in the reflectivity
dependence on frequency. It works, not very good though, for $Al$ but gives
considerably smaller value for $Au$ than that found from fitting $%
\varepsilon \left( \omega \right) $ in the infrared range. Probably such an
estimate was taken in \cite{Lam1,KRMM} for $Au$ where very small value $%
\omega _p=3.6\cdot 10^{15}\ s^{-1}$ was used.

The optimal fitting procedure is described in \cite{Gros}. The damping
frequency is evaluated first from the ratio $\left( 1-\varepsilon ^{\prime
}\right) /\varepsilon ^{\prime \prime }$ which depends linearly on frequency
in the Drude model

\begin{equation}
\label{omt}\frac{1-\varepsilon ^{\prime }}{\varepsilon ^{\prime \prime }}=%
\frac \omega {\omega _\tau }. 
\end{equation}

\noindent After that $\omega _p^2$ can be extracted from $1-\varepsilon
^{\prime }$ by linear fit. The results together with the statistical errors
are collected in Table \ref{1t} for those data in \cite{Zol} which include
the optical behavior of $Au$ and $Al$ in the infrared region. This table
clearly demonstrates that the Drude parameters depend significantly on the
sample which is used to measure the optical data. These samples contained
different densities of the defects (such as impurity atoms, vacancies,
dislocations, etc.) that influence their optical properties. In this sense
there are no universal material parameters. Reproducible parameters one can
get only for high-purity single-crystals. In this connection all the
attempts to use the handbook data for the Casimir force calculation can be
considered only as estimates and cannot claim on high precision.

Actually in any of the experiment we do not know the Drude parameters even
with 10\% accuracy. That is because the optical properties of evaporated or
spattered films which cover the bodies can be quite different from those of
bulk materials and depend on technological details of film preparation. It
is known, for example, that the film density is typically 0.7 from that of
the bulk material if it was not annealed. For the resistivity of spattered
and evaporated $Au$ \cite{Audat} the value $\rho _0=8.2\ \mu \Omega \cdot cm$
has been reported in contrast with the bulk resistivity $2.25\ \mu \Omega
\cdot cm$. If a metal is evaporated or spattered on a substrate, it has a
large number of defects. Relatively thick metallic films ($>100\ nm$) are
usually exist in polycrystalline form. Defects will reduce the concentration
of free electrons $n$ which defines the plasma frequency of the material.
They also will increase the damping frequency $\omega _\tau $ and
resistivity since the mean free path of electrons will shorten. To minimize
these undesirable in practical applications effects the films are usually
annealed at high temperature. In the experiments \cite{Lam1,MR,RM,RLM} it
was not reported were the bodies annealed or not but one can say definitely
that it was not done in the AFM experiments because the polysterene ball
cannot exist at the annealing temperature. Even more defects present in thin
films ($\leq 20-30\ nm$) which are usually amorphous. This explains why thin
films have very large resistivity in comparison with the bulk material.
Entries 1 and 4 for $Al$ in Table \ref{1t} correspond to the data for thick
film samples. They support our expectations that the plasma frequency for
films should be smaller and the resistivity larger than those for the bulk
material.

\section{\label{4} The upper limit}

Though we cannot use the handbook data to evaluate the force, one can
constrain it for a given experiment. This statement is based on the
observation that the force (\ref{shpl}) increases every time when $\omega _p$
increases or $\omega _\tau $ decreases. It has simple physical meaning: the
force becomes larger when the metal reflectivity increases. For us it is
important that any technological procedures will reduce $\omega _p$ and
increase $\omega _\tau $ for a given material. A perfect single-crystal will
have the largest plasma frequency and the smallest $\omega _\tau $ and these
parameters are well defined. One can use them to get the upper limit on the
Casimir force. The plasma frequency $\omega _p$ is defined by the
concentration of free electrons in the metal $n$ and their effective mass $%
m_e^{*}$

\begin{equation}
\label{omp}\omega _p=\sqrt{\frac{e^2n}{m_e^{*}\varepsilon _0}}, 
\end{equation}

\noindent where $e$ is the electron charge. For good metals, which we are
concerned, $m_e^{*}$ is close but larger than the electron mass. It will be
helpful for what follows to use Eq.\ (\ref{res}) and instead of the damping
frequency $\omega _\tau $ take the static resistivity $\rho \left( 0\right)
=\rho _0$ as a parameter. The later can be directly measured for any
material.

\subsection{Torsion pendulum experiment}

In the TP experiment \cite{Lam1} the quartz lens and plate were covered
first with $Cu$ of thickness $0.5\ \mu m$ and then with $Au$ of the same
thickness. The $Au$ layer is thick enough to be considered as infinite and $%
Cu$ will not play any role. We will find the upper limit on the electron
concentration if suppose that every $Au$ atom produce a free electron with
the mass $m_e^{*}=m_e$. Then for the $Au$ plasma frequency one finds $\omega
_p^{Au}=1.37\cdot 10^{16}\ s^{-1}$. The resistivity for crystalline gold is $%
\rho _0^{Au}=2.25\ \mu \Omega \cdot cm$. One can compare these parameters
with that given in Table \ref{1t} to make sure that they correspond to the
limit values. Substituting these parameters in (\ref{dfimag}) and
calculating the force according to (\ref{shpl}) one finds the upper limit on
the Casimir force $F^{max}\left( a\right) $ in the TP experiment.

To compare the upper limit on the force with the measured force $F^{exp}$, it
is more convenient to consider the residual force defined as\footnote{%
Note that Lamoreaux \cite{Lam1} used different definition of the residual
force $F^{exp}\left( a_i\right) -F_c^0\left( a_i\right) $.}

\begin{equation}
\label{resf}\Delta F(a_i)=F^{exp}\left( a_i\right) -F^{max}\left( a_i\right)
, 
\end{equation}

\noindent where $a_i$ are the separations for which the force has been
measured. Theory and experiment will be in agreement if $\Delta F$ will not
be positive within the experimental errors. The original experimental data
were presented for the lens curvature radius $R=11.3\ cm$ and the residual
force in this case is shown in Fig.\ \ref{2f}a. It clearly indicates the
presence of some unexplained force at the smallest separations. However,
later the author recognized \cite{Lam3} that he was working with aspheric
lens which had the curvature radius $R=12.5\pm 0.3\ cm$ in the place where
the force was measured. The correction was published in erratum \cite{Lam1}.
The points for $\Delta F(a_i)$ with the corrected $R$ are presented in Fig.\ 
\ref{2f}b. This time the prediction obviously does not contradict to the
experiment but dealing with the upper limit we cannot conclude that there is
an agreement, either.

The question about surface distortions in TP experiment has been raised in 
\cite{Most2}. Surfaces of the bodies have not been examined in \cite{Lam1}
but roughness of the order of $30-40\ nm$ is quite typical for a metallic
film on a polished substrate and correlates with the substrate roughness.
Quartz optics is used for near UV light and its surface has to be polished
with a precision at least $\lambda /10$, where $\lambda \sim 300\ nm$ is the
UV wavelength. It supports the value above which is routinely observed with
atomic force or tunnel microscope. According to (\ref{disflat}) the
short-scale stochastic distortions give only a few percent correction even
for the smallest separation $a=0.6\ \mu m$. Large-scale deviations seem
potentially more dangerous \cite{Most2} since the correction can be the
first order in $A/a$ especially if we take into account that the lens was
not spherical. The radius $r_{int}$ of the interaction area one can estimate
as $r_{int}\sim \sqrt{Ra}\sim 1000\ \mu m$. Therefore, only small area on
the lens takes part in the interaction. In this place the lens can be
represented as part of the parabolic surface

\begin{equation}
\label{parab}z=\frac{r^2}{2\left( R+\Delta R\right) }\approx \frac{r^2}{2R}-%
\frac 12\frac{r^2}{R^2}\Delta R, 
\end{equation}

\noindent where $\Delta R$ is the error in the curvature radius. Here the
second term describes the error in the plate-lens separation and it can be
taken as the distortion amplitude. This amplitude is maximal for $r=r_{int}$
and for the relative amplitude one has an estimate

\begin{equation}
\label{disam}\frac Aa\sim \frac 12\frac{\Delta R}R\simeq 1.2\cdot 10^{-2}. 
\end{equation}

\noindent This value is rather small and according to (\ref{disflat}) the
correction to the force will be less but comparable with the experimental
errors. Moreover, negligible role of the large-scale distortions is actually
an experimental fact. The region of the plate and sphere used for the force
measurement in \cite{Lam1} was varied by tilting the lens with the
adjustment screws and there was no evidence for any variation of the force
depending on the region used for the measurement.

The first attempt to evaluate the Casimir force was undertaken by the author
of TP experiment \cite{Lam1} who takes into account the first finite
conductivity correction but used very small $\omega _p$ for $Au$. Lamoreaux
was the first who recognized the necessity of more rigorous approach to the
force evaluation \cite{Lam2} and importance of thin films on the metallic
surfaces \cite{Lam3}. His numerical results were not quite good due to the
delicate problem with choice of $\omega _p$ which we discussed above. The
matter has been settled in \cite{LR1,LR2} with the result which coincide
with ours. However, our statement is that the calculated force represents
the upper limit but not the force itself. The reason is that evaporated $Au$
film will have smaller plasma frequency than $1.37\cdot 10^{16}\ s^{-1}$ due
to large number of defects in the film. To know the exact value of the
force, one has to measure the dielectric function of the bodies but not to
take it from a handbook.

\subsection{AFM experiment}

Let us discuss now the upper limit on the Casimir force for the AFM
experiment \cite{MR}. The plasma frequency for $Al$ can be restricted using (%
\ref{omp}) if one supposes that every atom produces 3 free electrons. It
gives $\omega _p^{Al}=2.40\cdot 10^{16}\ s^{-1}$ that coincide with the
largest value in Table\ \ref{1t}. The resistivity of perfect crystals is $%
\rho _0^{Al}=2.65\ \mu \Omega \cdot cm$. Since we successfully predicted the
plasma frequencies for the best samples of $Au$ and $Al$, the same way one
can use to estimate $\omega _p$ for $Au/Pd$. If each $Au$ atom gives one and 
$Pd$ atom gives not more than two free electrons, then $\omega
_p^{Au/Pd}=1.69\cdot 10^{16}\ s^{-1}$. This alloy is used in
microelectronics and resistivity of the bulk material is known to be $\rho
_0^{Au/Pd}\approx 30\ \mu \Omega \cdot cm$ \cite{Kriv} in accordance with
the statement that alloys have large Drude damping. These data allow to find
the upper limit on the force using (\ref{shpl}) with the functions $G_{1,2}$
defined in (\ref{defin2}).

Before comparing the upper limit with the measured force we have to discuss
a few additional aspects concerning the experiment. Real surface of the
bodies is always distorted and the distortions are especially important to
treat the data in \cite{MR}. The distortion statistics were analysed with
the atomic force microscope \cite{KRMM}. The force has to be averaged over
the distorted surfaces and we use for this the procedure developed in \cite
{KRMM}. The major distortions are the large separate crystals situated
irregularly on the surfaces with a typical lateral size of $200\ nm$. The
height of the highest and intermediate distortions is about $h_1=40\ nm$ and 
$h_2=20\ nm$, respectively. The homogeneous stochastic background of the
averaged height $h_0/2=5\ nm$ fills the surface between the major
distortions. The character of roughness on the plate and on the ball is
quite similar. The part of the surface occupied by distortions with the
height $h_1$, $h_2$, and $h_0/2$ was measured as ${\it v}_1=0.11$, ${\it v}%
_2=0.25$, and ${\it v}_0=0.64$, respectively. These values are the
probabilities for the corresponding distortion to appear. The body surface
is defined in such a way that averaging over distortions gives zero result.
Then the averaged force is the sum of local forces for all possible kinds of
distortions which face each other taken with the corresponding probabilities

\begin{equation}
\label{Fdis}F^{dist}\left( a\right) =\sum\limits_{i,j=0}^2{\it v}_i{\it v}%
_jF(a_{ij}), 
\end{equation}

\noindent where $a_{ij}$ are the local separations defined in \cite{KRMM}.
For us it will be important that the minimal local separation is $%
a_{11}=a-54.8\ nm$. This procedure seems quite reliable but, of course,
large distortions give the feeling of uncertainty. The upper limit on the
Casimir force has to be averaged with the corresponding roughness parameters
according to (\ref{Fdis}).

The raw force $F_m$ measured in the experiments consists of a few components 
\cite{MR} 
\begin{equation}
\label{fit}F_m=F_c(a_1+a_0)+F_e\left( a_1+a_0\right) +C\cdot \left(
a_1+a_0\right) . 
\end{equation}

\noindent Here $a_1$ is the separation from the voltage applied to the piezo
corrected to the cantilever deflection, $a_0$ is the parameter chosen in
such a way that $a=a_1+a_0$ is the absolute separation between bodies, the
first term in (\ref{fit}) is the Casimir force, the second term is the
electrostatic force corresponding to the measured contact potential $29\ mV$%
, the third term represents the linearly increasing coupling of the
scattered light into the photodiodes (see \cite{MR} for details). The
parameters $a_0$ and $C$ were determined at large separations, where $F_c$
is represented by the bare force (\ref{F0}). Then the Casimir force can be
extracted from the raw data with the help of (\ref{fit}). Similar way to
find the Casimir force was used in \cite{Lam1}. Of course, the
separation $a$ has to be defined as the distance between averaged surfaces
of $Au/Pd$ layers. However, the role of these layers have been
underestimated in \cite{MR,KRMM}. In \cite{MR} the Casimir force was found
from (\ref{fit}) but $a$ was interpreted as the absolute separation between $%
Al$ surfaces \cite{MR,KRMM}. Effectively the $Au/Pd$ layers were changed by $%
Al$ which has larger $\omega _p$ and, therefore, the force calculated
theoretically was overestimated.

It was indicated \cite{MR} that the thickness of $Au/Pd$ layer is less than $%
20\ nm$, that is why for calculations we use the conservative value $h=15\
nm $ . This change makes the force only larger. The experimental points from 
\cite{MR} (triangles) and theoretical upper limit on the force including the
roughness correction (solid line) are shown in Fig.\ \ref{3f} in the small
separations range $a<250\ nm$. If the top layer is changed by $Al$, it
enlarges the force on $15\ pN$ at the smallest separation $a=120\ nm$. It is
clear that the top layer definitely cannot be ignored in the force
evaluation. Variation of $\omega _p^{Al}$ on 10\% gives only $1\ pN$ change
in the force because of screening effect of the top layer. The same
variation in $\omega _p^{Au/Pd}$ changes the force on $2\ pN$. The
resistivity variation of the $Au/Pd$ layer on 30\% gives $1\ pN$ effect. At
larger separations all the effects become smaller.

We can see from Fig.\ \ref{3f} that the upper limit is smaller than the
force measured at small separations and the difference is significant. This
conclusion contradicts to that in \cite{KRMM}, where good agreement between
theory and experiment has been reported (dotted line) based on the detailed
theoretical analysis. We have already stressed the importance of the $Au/Pd$
layer but it is not the only reason of deviations. It comes also from poor
behavior of the finite conductivity correction used in \cite{KRMM} at small
separations $a$. The correction was based on a simple interpolating formula 
\cite{Cond3} for the force between two plates

\begin{equation}
\label{interp}F_c^{plates}(a)=F_c^{0\ plates}(a)\left( 1+\frac{11}3\frac c{%
a\omega _p}\right) ^{-\frac{16}{11}} 
\end{equation}

\noindent which is applicable in wider range of separations than the
expansion up to the second order (\ref{Fp}). It was used to calculate the
conductivity correction between sphere and plate. Applying the proximity
force theorem to (\ref{interp}) one gets

\begin{equation}
\label{sinterp}F(a)=3F_c^0(a)\int\limits_1^\infty \frac{dx}{x^4}\left( 1+%
\frac{11}{3x}\frac c{a\omega _p}\right) ^{-\frac{16}{11}}, 
\end{equation}

\noindent where the upper limit was moved to infinity since $a\ll R$. This
integral can be expressed via the Gauss hypergeometric function but in \cite
{KRMM} it was expanded in the series up to the fourth order

\begin{equation}
\label{Fp1}F(a)=F_c^0(a)\left[ 1-4\frac c{a\omega _p}+\frac{72}5\left( \frac 
c{a\omega _p}\right) ^2-\frac{152}3\left( \frac c{a\omega _p}\right) ^3+%
\frac{532}3\left( \frac c{a\omega _p}\right) ^4\right] 
\end{equation}

\noindent and the equations (\ref{Fdis}) and (\ref{Fp1}) were used to get
the theoretical prediction for the corrected Casimir force (dotted line in
Fig.\ \ref{3f}). We found that the interpolating curve which we got by
numerical intergation of (\ref{sinterp}) is very close to the exact force
evaluated according to (\ref{shpl}), (\ref{defin1}) with the parameters $%
\omega _p=1.88\cdot 10^{16}\ s^{-1}$, $\omega _\tau =0$ from \cite{KRMM} and 
$T=300^{\circ }K$ in all range of distances. The small difference between
the curves is the temperature effect which disappears when $T\rightarrow 0$.
However, the expansion (\ref{Fp1}) works very bad at $a<100\ nm$. Although
the separation in the experiment exceeds $100\ nm$, the local distance in (%
\ref{Fdis}) can be as small as $65\ nm$, where (\ref{Fp1}) is absolutely
unacceptable. The same is true if one uses the expansion up to the fourth
order found in \cite{Cond4} directly form (\ref{shpl}). The dashed line in
Fig.\ \ref{3f} shows the force calculated according to (\ref{Fdis}), (\ref
{sinterp}) with the parameters above. The divergence of the solid and dashed
curves is the effect of the top layer and different parameters used for $Al$%
. The larger $\omega _p^{Al}$ which we are using for the upper limit is
partly compensated by the top layer and that is why the dashed curve lies
not too far from the solid one.

\subsection{Improved AFM experiment}

Very important progress has been achieved in \cite{RLM} where controlled
metal evaporation and smaller thickness of $Au/Pd$ layer ($h=8\ nm$ instead
of $20\ nm$) allowed to reduce the surface roughness to the level when the
correction to the force becomes practically unimportant. Also the contact
potential has been considerably reduced and the parameter $a_0$ defining the
absolute separation of the surfaces has been found in independent
electrostatic measurements. Interaction between metallized ball and
corrugated plate has been probed in \cite{RM}. It is not a subject of our
consideration here. The data for flat plate and sphere in this work are
actually in very good agreement with that given in \cite{RLM} and we will
not discuss them specially. The roughness parameters have been reduced to
the following \cite{RLM}: $h_1=14\ nm,\ {\it v}_1=0.05;\ h_2=7\ nm,\ {\it v}%
_2=0.11;\ h_0/2=2\ nm,\ {\it v}_0=0.84$. Unfortunately, the unjustified
assumption that the force is insensitive to the presence of $Au/Pd$ layer
has been inserted in the procedure of the Casimir force extraction from the
raw data and the following relation has been used instead of (\ref{fit}) 
\begin{equation}
\label{fit1}F_m=F_c(a+2h)+F_e\left( a\right) +Ca. 
\end{equation}

\noindent For this reason we cannot directly use the data in \cite{RLM} to
compare with the theoretical prediction. It becomes obvious if we plot in
the same figure the measured force in the experiments \cite{MR} and \cite
{RLM} (see Fig.\ \ref{4f}). One would expect that for thicker $Au/Pd$ layer
the force has to be smaller, but the actual relation is opposite and the
difference is large. Fortunately, it is easy to restore the right data. In 
\cite{RLM} $a_0$ was found from an independent electrostatic measurement and
the constant $C$ was determined at large separations when the shift on $%
2h=16\ nm$ in $F_c$ argument was practically unimportant. The measured
Casimir force $F_{c-m}$ was expressed as

\begin{equation}
\label{fit2}F_{c-m}(a+2h)=F_m(a)-F_e(a)-Ca, 
\end{equation}

\noindent but the points in Fig.\ \ref{4f} taken from \cite{RLM} represent
the force as a function of true separation. Therefore, the force presented
in the figure was calculated as

\begin{equation}
\label{fit3}F_{c-m}(a)=F_m(a-2h)-F_e(a-2h)-C(a-2h). 
\end{equation}

\noindent The $Au/Pd$ layer certainly cannot be ignored and the right
expression for the measured Casimir force must be

\begin{equation}
\label{fit4}F_{c-m}(a)=F_m(a)-F_e(a)-Ca. 
\end{equation}

\noindent It is obvious that to restore the right data one has to shift the
open squares in Fig.\ \ref{4f} on $2h=16\ nm$ to larger separations. After
this shift a good agreement between two different experiments is reached. To
be absolutely sure that the right transformation was done we have tried to
reproduce the measured force directly from the raw data presented in \cite
{MR} and \cite{RLM}, where the procedure has been described in details. The
data were available only for one scan and for this reason our calculations
had restricted precision, but it was enough to make a conclusion on
reproducibility. To check the procedure, we successfully reproduced the
force from the raw data in \cite{MR}. The force found from the raw data in 
\cite{RLM} according to (\ref{fit4}) agrees much better with the shifted
points than with the ones presented in \cite{RLM}. These detailed
explanations are given not only to answer the criticism of our preprint \cite
{SL} but also because of great importance of the conclusion. It is stated in 
\cite{Yuk,KMM} that the points have to be shifted to smaller separations.
Even ignoring the arguments above it is obvious from Fig.\ \ref{4f} that
such a shift would give drastic disagreement between two experiments made by
the same method.

The upper limit on the Casimir force in conditions of the experiment \cite
{RLM} one can find exactly as was explained above. The only difference is
the other set of roughness parameters, but in this case the roughness
correction is on the level of experimental errors. The experimental points
from \cite{RLM} shifted on $16\ nm$ as was discussed above and the
corresponding upper limit (solid line) are presented in Fig.\ \ref{5f}.
Again we can see that the upper limit is smaller than the measured force and
deviation increases at smaller separations. Moreover, even if we replace the 
$8\ nm$ thick top layer by $Al$ with the maximal plasma frequency $\omega
_p=2.40\cdot 10^{16}\ s^{-1}$, this disagreement will not disappear as shows
the dashed line. The residual force defined according (\ref{resf}) in the
experiments \cite{MR} (triangles) and \cite{RLM} (open squares) is shown in
Fig.\ \ref{6f}. It clearly demonstrates the presence of some unexplained
attractive force which is decreasing rapidly when the distance between
bodies increased. The points from two different experiments are in
reasonable agreement with each other that means that the residual force is
reproducible. The residual force becomes larger if we deviate the parameters
from their limit values but the agreement between two experiments is not
broken.

\section{\label{5}Discussion}

Let us discuss now possible reasons for disagreement between experiment and
theoretical expectations. As was mentioned above the main problem is the
values of the material parameters which can significantly deviate from their
handbook values for evaporated or spattered metallic films. The idea of this
paper was to find the upper limit on the force instead of the force itself.
It allowed to use only well defined parameters of perfect single-crystal
materials. We took the largest values for the plasma frequencies and the
smallest ones for the resistivities. Any possible deviation from these
values will make the force only smaller and disagreement between theory and
experiment will be larger.

Some doubts were raised \cite{KMM} about the possibility to describe the
thin top layer by a dielectric function which depends only on frequency. It
was stated that the spatial dispersion can be important for thin films
because the distance traveled by electron during one period of the field can
be larger than the film thickness

\begin{equation}
\label{sdisp}\frac{{\it v}_F}\omega >h, 
\end{equation}

\noindent where ${\it v}_F$ is the velocity of the electron on the Fermi
surface. This dimensional effect is really exist (see, for example, \cite{CM}%
) but it is difficult for observation at room temperature. The reason is
that for thin metallic films the mean free path for electrons is very short (%
$<100\ \AA $) because of large concentration of the defects. Typically the
resistivity of very thing films is on the level of $100\ \mu \Omega \cdot cm$%
. Then for $\omega _p\sim 10^{16}\ s^{-1}$ from (\ref{res}) one finds $%
\omega _\tau \sim 10^{15}\ s^{-1}$. The mean free path is estimated as

\begin{equation}
\label{mfp}l=\frac{{\it v}_F}{\omega _\tau }\sim 10\ \AA 
\end{equation}

\noindent that is smaller than the used $Au/Pd$ film thickness and,
therefore, the spatial dispersion can be neglected. That is why the standard
dependence for the dielectric function $\varepsilon \left( \omega \right) $\
is widely used in optics of metals up to the film thickness in a few
nanometers when quantum effects become involved. For the same reason one can
neglect the anomalous skin effect for evaporated (spattered) films even for
thick ones. Extremely high resistivity $2000\ \mu \Omega \cdot cm$ for $60\
nm$ thick $Au/Pd$ film has been reported in \cite{KWB}. However, the authors
themselves stress that the resistance of the film is, to all appearance,
dominated by grain boundaries but optical properties of the film are quite
usual. This example shows once more that the details of the spattering
technology cannot be ignored.

In \cite{LR1} uncertainty was expressed about applicability of the proximity
force theorem. At the moment there is no any work where the force between
sphere and plate was calculated from ''the first principles''. There are
some heuristic approaches \cite{book} allowing to calculate nonadditive
Casimir force which agree well with the result found by using PFT (see
additional discussion and references in \cite{KMM}). The PFT states that the
main contribution to the force can be found by adding the contributions of
various distances as if they were independent and it is applicable to
nonadditive forces. An example is the electrostatic force which is
nonadditive because the surface charge density is nonuniform for curved
surfaces. One can check that the PFT gives in this case the correct result
with the accuracy $\sim a/R$. For the discussed experiments the correction
is very small ($a/R\sim 0.001$). Even if this term appears with a large
coefficient, say $\sim 10$, the correction will be only on the level of the
experimental errors.

In the AFM experiments the electrostatic attraction between bodies because
of contact potential was carefully taken into account. Of course, the
aluminum surfaces were partly oxidized and electrons could be trapped in the
oxide. These charges can be potentially dangerous if the $Au/Pd$ film is not
continuous. In this case the trapped charges and their images in underlying
aluminum will be the source of the dipole field. Then an additional force
can arise as a result of dipole-dipole interaction. However, it is difficult
to make a reliable estimate for this effect because we do not know the
concentration of trapped charges and the size of islands in $Au/Pd$ layer or
even do discontinuities exist at all for the used layer thickness (it
depends on details of the covering procedure). In this connection to make
the experiment absolutely clear, it is preferable to use $Au$ instead of $Al$
metallization because its non-reactive surface has strong advantage over $Al$%
. It excludes also additional uncertainties connected with $Au/Pd$ layer.
One can use as well silver or copper but they are not as inert as gold. It
is difficult to measure the dielectric function at the wavelengths larger
than $30\ \mu m$ but this range gives an important contribution to the
dispersion relation. That is why the material behavior in this range has to
be predictable. One can say definitely that the materials of platinum group
cannot be used since they are not described by the Drude dielectric function
at low frequencies.

One can speculate that the observed discrepancy is explained by a new Yukawa
force mediated by a light scalar boson. Then interaction of two atoms is
described by the Yukawa potential

\begin{equation}
\label{VY}V_Y\left( r\right) =-\alpha N_1N_2\frac{\hbar c}r\exp (-r/\lambda
), 
\end{equation}

\noindent where $\alpha $ is a dimensionless interaction constant, $\lambda $
is the Compton wavelength of a particle responsible for the interaction, and 
$N_{1,2}$ is the number of nucleons in atoms of the interacting bodies. An
additional advantage of $Au$ metallization is higher density of the bodies
coating. In this case the Yukawa force will be enlarged roughly by the
factor $\left( \rho _{Au}/\rho _{Al}\right) ^2\approx 50$, where $\rho
_{Au,Al}$ are the material densities. If the observed discrepancy has
relation with the Yukawa interaction, the AFM experiment with $Au$
metallization of the bodies will definitely reveal this new force even
without detailed knowledge of optical properties of the metallization.

\section{Conclusion}

We have analysed the results of recent precise measurements of the Casimir
force using the Lifshitz theory to evaluate the force. Layered structure of
the bodies coating was taken into account in the frame of the Lifshitz
approach. It was stressed that the force cannot be predicted with necessary
accuracy if there is no detailed information on the dielectric function of
the bodies coating. Fortunately, all the used materials ($Au$, $Al$ and $%
Au_{0.6}Pd_{0.4}$) are well described in terms of Drude parameters $\omega
_p $ and $\omega _\tau $ in the infrared range which dominates in the
dispersion relation for the dielectric function $\varepsilon \left( i\zeta
\right) $. It was noted that one can find the upper limit on the Casimir
force that realized for perfect single-crystal materials for which
electrical and optical properties are well defined. The surface roughness
and linear in temperature corrections were taken into consideration. It was
shown that the upper limit on the Casimir force does not contradict to the
result of the torsion pendulum experiment \cite{Lam1}. The main conclusion
of the paper is that the upper limit is smaller than the observed force in
the AFM experiments and the difference far exceeds experimental errors and
theoretical uncertainties for small separations between bodies. The simplest
modification of the experiment is proposed allowing to reveal origin of the
discrepancy.

\begin{figure}
\caption{This figure demonstrates validity of the Drude approximation
for $Al$ (triangles) and $Au$ (circles) in the infrared range. The 
resistivity does not depend on frequency (left axis). Solid lines
(right axis) show that $\varepsilon ^{\prime \prime }\left(\omega \right) $
depends on $\omega $ according to (\protect \ref{Drudr}) with the 
parameters given in Table\ \protect\ref{1t} (entries 2).}
\label{1f}
\end{figure}

\begin{figure}
\caption{The residual force (\protect\ref{resf}) as a function of sphere-plate separation
$a$ for TP experiment \protect \cite{Lam1}. For the original value of sphere 
radius $R=11.3\ cm$ the points of closest approach in (a) demonstrate presence of some 
unexplained force. With the corrected value $R=12.5\ cm$ \protect \cite{Lam3} 
the residual force is shown in (b). In this case there is no contradiction 
between theory and experiment.}
\label{2f}
\end{figure}

\begin{figure}
\caption{The Casimir force measured in the AFM experiment \protect \cite{MR} 
(triangles). The solid line represents the upper limit on the force. 
The dotted line is taken from \protect \cite{KRMM} where the $Au/Pd$ layer 
was ignored and expansion (\protect\ref{Fp1})
used for the finite conductivity correction. This expansion fails at small 
separations and Eq.\ (\protect\ref{sinterp}) has to be used instead. The
result is represented by the 
dashed line. The difference between the solid and dashed lines is due 
to $Au/Pd$ layer.}
\label{3f}
\end{figure}

\begin{figure}
\caption{Comparison of the results of two AFM experiments. The data from 
\protect \cite{MR} are marked by solid triangles and the data from 
\protect \cite{RLM} presented as the open 
squares. The obvious contradiction of two experiments made with the same 
technique is connected with unjustified assumption of the transparency of 
$Au/Pd$ layer in \protect \cite{RLM}. The right data can be restored by 
simple shift all the open squares on $16\ nm$ to larger separations 
(see explanations in the text) and after that the experiments will agree 
with each other.}
\label{4f}
\end{figure}

\begin{figure}
\caption{The data from \protect \cite{RLM} shifted on $16\ nm$ (open squares) 
and the upper limit on the Casimir force (solid line). The dashed line 
represents the
force for the case when the $8\ nm$ thick $Au/Pd$ top layer is changed by 
$Al$ with the maximal plasma frequency $\omega_p=2.4 \cdot 10^{16}\ s^{-1}$.}
\label{5f}
\end{figure}

\begin{figure}
\caption{The residual force defined as (\protect \ref{resf}) for the two AFM experiments.
The data from \protect \cite{MR} and \protect \cite{RLM} are presented as 
solid triangles and open
squares, respectively. The figure demonstrates presence of some unexplained force 
which decreases rapidly with separation increase. The residual force is reproduced 
for both of the experiments.}
\label{6f}
\end{figure}

\begin{table}   \begin{tabular}{|c|c|c|c|} \hline
     {\bf Al}  &  $\omega_p \cdot 10^{-16} $  &  $\omega_{\tau} %
     \cdot 10^{-13} $  &  $\rho_0 \ \mu \Omega \cdot cm$ \\ \hline
     $1^*$       & $1.54\pm 0.01$ & $15.5\pm 0.6$  & 7.39 \\
     2       & $2.235\pm 0.001$ & $12.49\pm 0.01$ & 2.83 \\
     3       & $2.43\pm 0.05$ & $14.4\pm 0.7$ & 2.76 \\ 
     $4^*$       & $1.63\pm 0.03$ & $18.2\pm 0.7$ & 7.74 \\ \hline
      {\bf Au}  &  $\omega_p \cdot 10^{-16} $  &  $\omega_{\tau} %
      \cdot 10^{-13} $  &  $\rho_0 \ \mu \Omega \cdot cm$ \\ \hline
     1       & $1.280\pm 0.001$ & $3.29\pm 0.05$  & 2.27 \\
     2       & $1.3720\pm 0.0006$ & $4.060\pm 0.002$ & 2.44\\
     3       & $1.34\pm 0.02$ & $7.08\pm 0.18$ & 4.46\\ 
     4       & $1.0513\pm 0.0007$ & $6.24\pm 0.21$ & 6.40 \\ \hline
  \end{tabular}
  \caption{Parameters of the Drude dielectric function 
   (\protect\ref{Drudr}) for $Al$ and $Au$ found by fitting the data 
   \protect \cite{Zol} of different measurements in the infrared range 
   ($\lambda > 2 \ \mu m$). The statistical errors of fitting are indicated. 
   The resistivity was calculated via $\omega_p$ and $\omega_{\tau}$ according
   to (\protect\ref{res}). 
   The stars in the first column mark the data for film samples.}
\label{1t}
\end{table}


\begin{references}
\bibitem{Casimir}  H.B.G. Casimir, Koninkl. Ned. Akad. Wetenschap. Proc. 
{\bf 51}, 793 (1948).

\bibitem{book}  V.M. Mostepanenko and N.N. Trunov, {\it The Casimir effect
and its applications} (Clarendon Press, Oxford, 1997).

\bibitem{Lif}  E.M. Lifshitz, Sov. Phys. JETP {\bf 2}, 73 (1956).

\bibitem{LP}  E.M. Lifshitz and L.P. Pitaevskii, {\it Statistical Physics,
Part 2}, (Pergamon Press, Oxford, 1980).

\bibitem{Kuz}  V.A. Kuz'min, I.I. Tkachev, and M.E. Shaposhnikov, JETP
Letters (USA) {\bf 36}, 59 (1982).

\bibitem{Most1}  V.M. Mostepanenko and I.Yu. Sokolov, Phys. Lett. {\bf A 125}
, 405 (1987).

\bibitem{Long}  J.C. Long, H.W. Chan, and J.C. Price, Nucl. Phys. {\bf B 539}%
, 23, (1999).

\bibitem{Fisch}  E. Fischbach and C. Talmadge, {\it The Search for
Non-Newtonian Gravity} (AIP Press/Springer-Verlag, New York, 1999).

\bibitem{Lam1}  S.K. Lamoreaux, Phys. Rev. Lett. {\bf 78}, 5 (1997); {\bf 81}%
, 5475(E) (1998).

\bibitem{MR}  U. Mohideen and A. Roy, Phys. Rev. Lett. {\bf 81} , 4549
(1998); e-print physics/9805032.

\bibitem{RM}  A. Roy and U. Mohideen, Phys. Rev. Lett. {\bf 82} , 4380
(1999).

\bibitem{RLM}  A. Roy, C. -Y. Lin, and U. Mohideen, Phys. Rev. {\bf D 60},
111101 (1999); e-print quant-ph/9906062.

\bibitem{PFT}  J.Blocki, J. Randrup, W.J. Swiatecki, and C.F. Tsang, Ann.
Phys. (N.Y.) {\bf 105},427 (1977).

\bibitem{Cond1}  I. E. Dzyaloshinskii, E. M. Lifshitz, L. P. Pitaevskii,
Sov. Phys. Uspekhi {\bf 4} (1961) 153.

\bibitem{Cond2}  C. M. Hargreaves, Proc. Kon. Nederl. Acad. Wet. {\bf B68}
(1965) 231; J. Schwinger, L.L. DeRaad, Jr., and K.A. Milton, Ann. Phys.
(N.Y.) {\bf 115}, 1 (1978).

\bibitem{Cond3}  V.M. Mostepanenko and N.N. Trunov, Sov. J. Nucl. Phys.
(USA) {\bf 42}, 812 (1985); V. B. Bezerra, G. L. Klimchitskaya, C. Romero,
Mod. Phys. Lett. {\bf A 12} (1997) 2623.

\bibitem{Temp}  J. Mehra, Physica {\bf 37}, 145 (1967); L.S. Brown and G.J.
Maclay, Phys. Rev. {\bf 184}, 1272 (1969); J. Schwinger, L.L. DeRaad, Jr.,
and K.A. Milton, Ann. Phys. (N.Y.) {\bf 115}, 1 (1978).

\bibitem{Dist}  M. Bordag, G.L. Klimchitskaya, and V.M. Mostepanenko, Int.
J. Mod. Phys {\bf A 10}, 2661 (1995).

\bibitem{KRMM}  G. L. Klimchitskaya, A. Roy, U. Mohideen, and V. M.
Mostepanenko, Phys. Rev. {\bf A 60}, 3487 (1999); e-print quant-ph/9906033.

\bibitem{Lam2}  S.K. Lamoreaux, Phys. Rev. {\bf A59} R3149 (1999).

\bibitem{LR1}  A. Lambrecht and S. Reynaud, Europ. Phys. J. D {\bf 8} 309
(2000); e-print quant-ph/9907105.

\bibitem{LR2}  A. Lambrecht and S. Reynaud, Phys. Rev. Lett., to appear;
e-print quant-ph/9912085.

\bibitem{SL}  V.B. Svetovoy and M.V. Lokhanin, e-print quant-ph/0001010.

\bibitem{Cond4}  V.B. Bezerra, G.L. Klimchitskaya, V.M. Mostepanenko,
e-print quant-ph/9912090.

\bibitem{SLtem}  V.B. Svetovoy and M.V. Lokhanin, e-print quant-ph/0004004.

\bibitem{Lam3}  S.K. Lamoreaux, e-print quant-ph/9907076.

\bibitem{KMM}  G. L. Klimchitskaya, U. Mohideen, and V. M. Mostepanenko,
e-print quant-ph/0003091.

\bibitem{Gros}  P. Grosse, {\it Free electrons in solids}, (Springer-Verlag,
Berlin,1979).

\bibitem{Pers}  B.N.J. Persson, J.E. Demuth, Phys. Rev. {\bf B 31}, 1856
(1985).

\bibitem{Zol}  V.M. Zolotarev, V.N. Morozov, and E.V. Smirnov, {\it Optical
constants of natural and technical mediums }(Chimiya, Leningrad,1984).

\bibitem{hbook}  {\it Handbook of Optical Constants of Solids,} E.D. Palik
ed. (Academic Press, New York 1995).

\bibitem{Audat}  L. Lechevallier, G. Richon, J. Le Bas, and M. Bernole,
Vacuum {\bf 41}, 1218 (1990).

\bibitem{Most2}  M. Bordag, B. Geyer, G.L. Klimchitskaya, and V.M.
Mostepanenko, Phys. Rev. {\bf D 58}, 075003 (1998).

\bibitem{Kriv}  {\it Materials in mechanical engineering and automation}. 
{\it Handbook. }Edited by{\it \ } Yu.M. Pyatin, 2nd ed. (Mashinostroenie,
Moskva, 1982).

\bibitem{Yuk}  M. Bordag, B. Geyer, G. L. Klimchitskaya, and V. M.
Mostepanenko, e-print hep-ph/0003011.

\bibitem{CM}  D.S. Campbell and A.K. Morley, Rep. Prog. Phys. {\bf 34, }283
(1971).

\bibitem{KWB}  J.R. Kirtley, S. Washburn, and M.J. Brady, Phys. Rev. Lett. 
{\bf 60}, 1546 (1988).
\end{references}
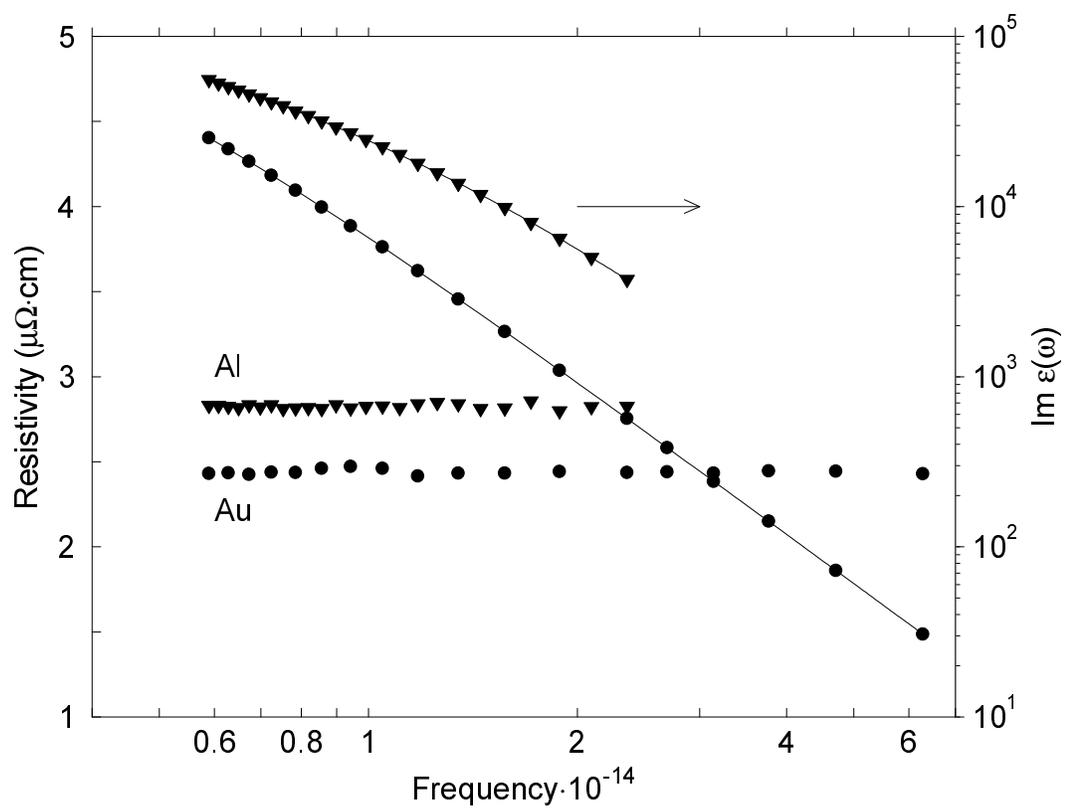
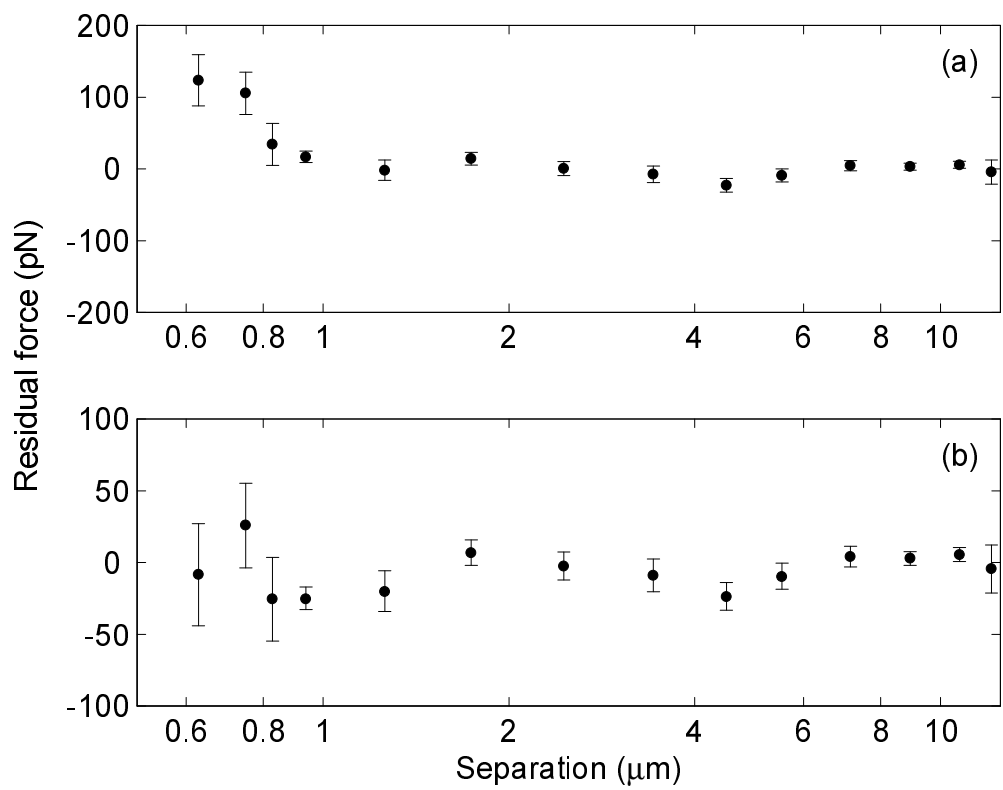
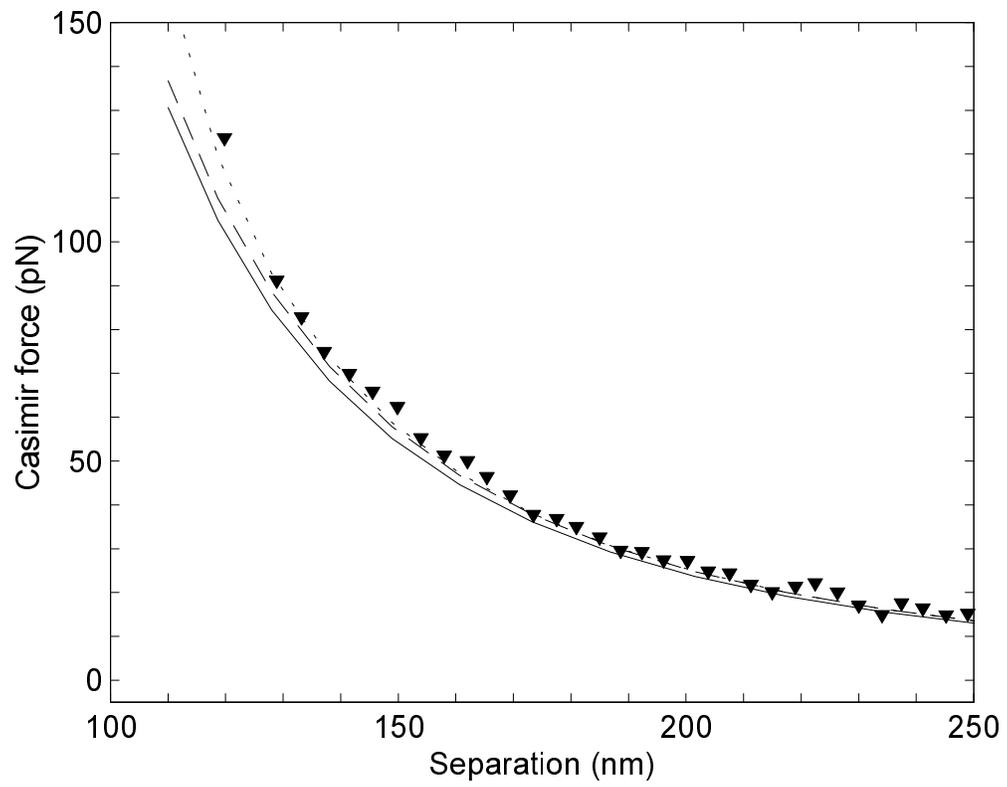
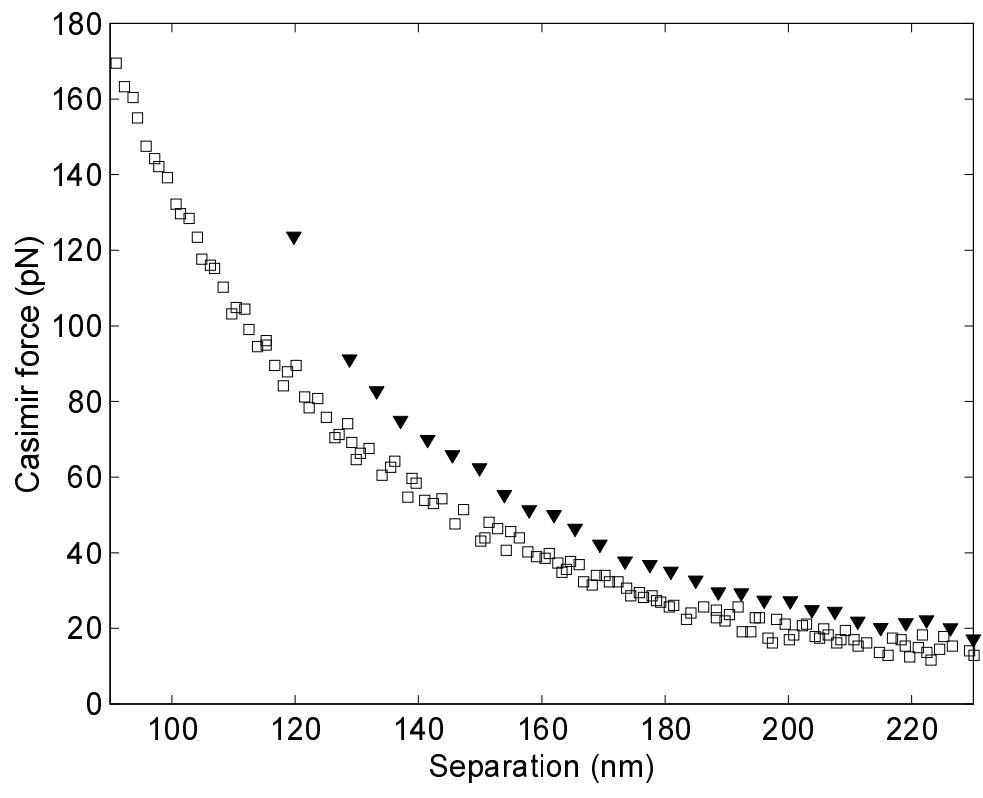
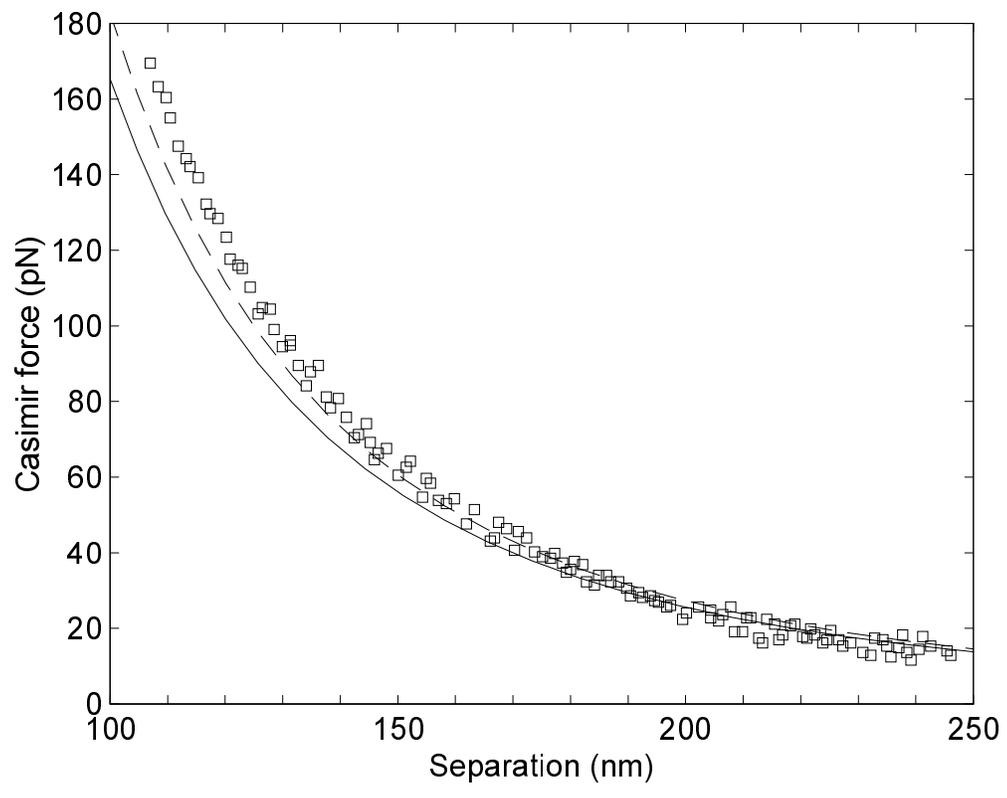
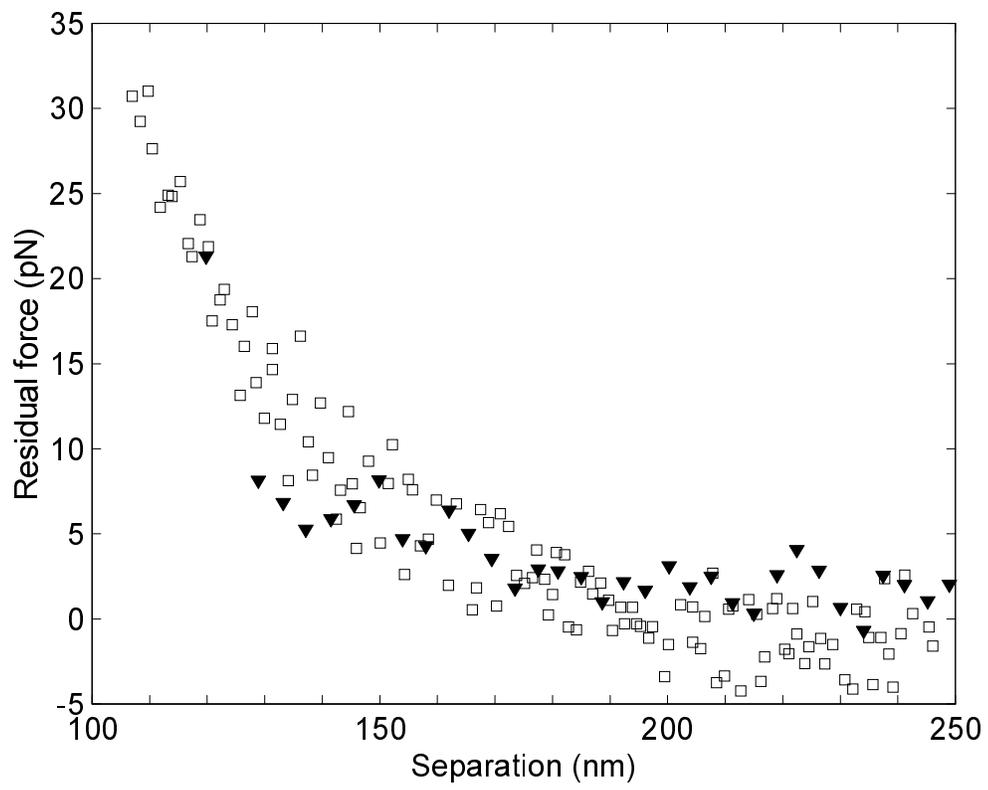
\end{document}